# Distributed Augmented Reality with 3D Lung Dynamics
## - A Planning Tool Concept -

Felix G. Hamza-Lup, Anand P. Santhanam, Celina Imielinska, Sanford Meeks and Jannick P. Rolland

*Abstract* - **Augmented Reality (AR) systems add visual information to the world by using advanced display techniques. The advances in miniaturization and reduced costs make some of these systems feasible for applications in a wide set of fields. We present a potential component of the cyber infrastructure for the operating room of the future; a distributed AR based software-hardware system that allows real-time visualization of 3D lung dynamics superimposed directly on the patient's body. Several emergency events (e.g. closed and tension pneumothorax) and surgical procedures related to the lung (e.g. lung transplantation, lung volume reduction surgery, surgical treatment of lung infections, lung cancer surgery) could benefit from the proposed prototype.**

## I. INTRODUCTION

Augmented Reality (AR) allows the development of promising tools in several domains from design and manufacturing [1] to medical applications [2, 3]. We present a potential component of the cyber infrastructure for the operating room of the future: a distributed AR based software-hardware system that allows real-time visualization of 3D lung dynamics superimposed directly on the patient's body. The application domain for the proposed cyber infrastructure includes training students on clinical procedures (e.g intubation) and planning clinical interventions. Specifically, pre, intra, and post-operative assessments for emergency events (e.g. closed and tension pneumothorax) and pre and post-operative assessments for surgical procedures related to the lung (e.g. lung transplantation, lung volume reduction surgery, surgical treatment of lung infections, lung cancer surgery) could benefit from the proposed visualization tool. The tool also facilitates experts' interactions, especially during quick-response conditions such as medical emergencies in geographically inaccessible locations [4]. In this work, an AR based visualization system is integrated with a Human Patient Simulator (HPS). This integrated prototype provides a simulation and training test-bed that most closely resembles the clinical end application.

This work exemplifies advanced medical visualization paradigms where known methodologies in different technical fields are combined and tailored to best serve the requirements of the application. Specifically, we use the methods discussed in [5, 6] to model the real-time 3D lung dynamics and its biomathematics, respectively.

An early integration of the real-time 3D lung dynamics with a distributed AR based framework was presented in [7], where the real-time 3D lung dynamics were superimposed on the HPS. In this earlier integration, we demonstrated the possibility for remotely located clinical technicians to view the 3D lung dynamics of a patient in real-time. We extend our previous work by concentrating on the integration of the hardware and software for the development of a prototype that could be part of the operating rooms in the near future.

The paper is structured as follows. Section II presents the AR planning tool concept followed by related work on modeling 3D lung dynamics in Section III. Section IV summarizes the hardware components involved. Modeling 3D lung dynamics and the data distribution scheme are presented in Section V. Results and preliminary assessment, Section VI, is followed by discussions, conclusions, and future work in Sections VII and VIII.

## II. PLANNING TOOL CONCEPT

The distributed interactive tool aimed at surgical planning allows visualization of virtual 3D deformable lung models overlaid on the patient body in an effort to improve the medical planning process.

The planner and other medical personnel remotely located may be involved in the surgical planning procedure and may visualize the 3D lung model while seeing and interacting with each other in a natural way (Figure 1a). Moreover, they are able to participate in the planning procedure by pointing and drawing diagrams in correspondence with the 3D lung model. Figure 1b shows the super-imposition concept: the virtual 3D model of the lung overlaid on the patient's thoracic cage.

Felix G. Hamza-Lup, Ph.D. is Assistant Professor of Computer Science at the School of Computing, Armstrong Atlantic State University, Savannah. *Email: felix@cs.armstrong.edu.*

Anand P. Santhanam, School of Computer Science at the Univ. of Central Florida (UCF), Orlando.

Celina Imielinska, Ph.D. is Associate Research Scientist, Department of Bio-informatics at the Columbia University, New York.

Sanford L. Meeks, Ph.D. is Director of Radiation Physics at M. D. Anderson Cancer Center Orlando.

Jannick P. Rolland, Ph..D. is Associate Professor of Optics at the College of Optics and Photonics (UCF).

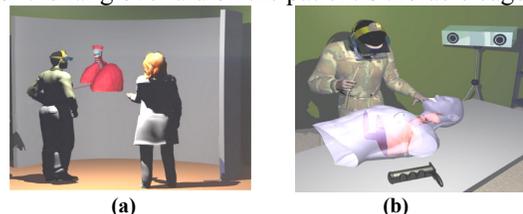

Figure 1. (a) Surgical planning personnel interacting with 3D models and (b) User's view of the virtual 3D model of the lungs as he/she interacts with participants in (a).

The remote medical personnel interacting with the (local) planner can see the changes in the lung's behavior and will adopt different planning choices.

Such an advanced planning tool has the potential to:

- Involve local and remote planning personnel in the surgical planning procedure, opening new ways of collaboration and interaction.
- The local planner can actually "see" the anatomical model (patient specific data) superimposed on the patient and improve the surgical plan.

## III.   RELATED WORK

With these concepts in mind, we provide a brief review of techniques and algorithms that support rendering of dynamic 3D lung models and an important issue in distributed simulations, the dynamic shared state.

### A. DEFORMABLE LUNG MODELS

Early attempts to model the lung deformation were based on physiology and clinical measurements [8]. The physically-based deformation of the lungs as a linearized model was proposed by Promayon [9] followed by a Finite Element Model (FEM) based deformation for modeling pneumothorax-like conditions [10]. An early functional FEM model for the lung tissue constituents (i.e., parenchyma, bronchiole and alveoli) of lungs aimed at analyzing the anatomical functions of the lung during breathing [11]. The computational complexity of the approach was reduced by modeling only the bronchioles and the air-flow inside them [12].

Non-physically-based methods for lung deformation combined non-uniform rational B-spline surfaces based on data from patients' Computed Tomography scans [13].

### B. DISTRIBUTED INTERACTIVE SIMULATIONS

In distributed interactive simulations the interactions and information exchanges generate a state referred to as the *dynamic shared state,* which has to remain consistent for all participants at all sites. The interactive and dynamic nature of a distributed simulation is constrained by the communication latency, as well as by the complex visualization and rendering systems latency.

A number of consistency maintenance techniques have been employed in distributed environments and the research efforts can be grouped in four categories: communication protocol optimization, virtual space management, human perceptual limitation [14] and system architecture [15].

AR systems were proposed in the mid '90s as tools to assist different fields: medicine [16], complex assembly labeling [17], and construction labeling [18]. With advances in computer graphics, tracking systems, and 3D displays, the research community has shifted attention to distributed collaborative environments that use extensively the AR paradigm [19].

## IV.   HARDWARE COMPONENTS

Our system integrates a 3D visualization device with an optical tracking system and a Linux-based PC. With the exception of the 3D visualization device, the prototype was integrated using commercially available hardware components.

### A. VISUALIZATION DEVICE

To see the 3D model, the planner and the other users wear lightweight head-mounted displays (HMDs) [20] (Figure 2a) .

A key component in the reduction of weight in the optical design for the HMD is the optimal integration of diffractive and plastic optics as well as various emerging optical materials that may be used to compensate for optical aberrations responsible for image degradation. Optics as light as 6 grams per eye have been achieved for horizontal field of views as large as 70 degrees.  The current projection resolution of the HMD is 600x800 pixels.

### B. TRACKING SYSTEM

To superimpose the 3D virtual models at the correct location with respect to the patient, we need to track, in real–time, the relative position of the visualization device or planner's head and the patient's thoracic cage [21].

The pose (i.e., position/orientation) of the planner's head, and the patient's thoracic cage are determined using an optical tracking system (i.e., Polaris™ Northern Digital) and two custom built tracking probes [22] illustrated in Figure2_ab.

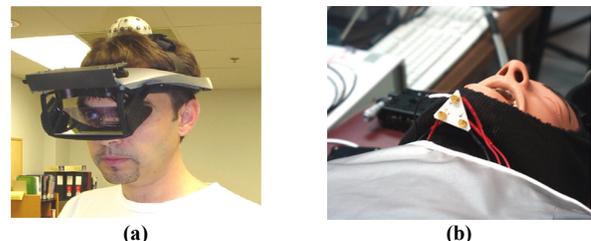

Figure 2. Collections of LED (i.e., tracking probes): (a) Optical see-through HMD developed through an interdisciplinary research effort at ODALab UCF (Courtesy- NVIS inc. for the opto-mechanical design) and custom designed semispherical head tracking probe, (b) patient tracking probe.

A key step in the accurate registration of the virtual lungs over the patient is the calibration procedure. In this step, the patient is positioned in the intubation posture. Magnetic Resonance Imaging (MRI) data of the relative position of the mandible, selected landmarks with respect to the larynx, trachea, and the clavicle were collected in this posture. The inclusion of mandible, larynx and trachea in the above step allows us to track the position of the intubation tube (through the mouth and upper-airway), for medical procedures that require such support. The lung can be registered to the mandible and the upper chest landmarks (e.g. clavicle). From that point on, the lungs are kept in registration with the patient, based on the location of the upper chest landmarks, together with the position of the user.

The tracking system, based on two tracking probes: one on the HMD to determine the planner's head position and orientation, and a second on the chin of the HPS to determine its location. The tracking data obtained is currently updated at

40Hz. The tracking working volume is a cone having 1.5 meters in height and 0.5 meters in radius.

To superimpose the 3D virtual model of the lung on the patient we used a least-squares pose estimation algorithm detailed in [23]. The predefined set of markers on the head tracking probe are uniformly distributed on a hemisphere to enable 360 degrees tracking while the user tilts his/her head.

## V. SOFTWARE COMPONENTS

Further we detail the 3D lung dynamics modeling approach and the data distribution scheme.

### A. THE DEFORMABLE LUNG MODEL

The virtual 3D deformable lung model contains two components. For the first component, we parameterized the Pressure-Volume (PV) i.e. the relation between the lung volume and the trans-pulmonary pressure [24]. The relation represented in terms of a set of control constants and basis functions, allowed us to model subject-specific breathing patterns and their variations. Variations in breathing may be caused by abnormalities such as pneumothorax, tumors or dyspnea. These variations lead to changes in the PV curve of the patient and can be modeled by controlled modulations of the basis functions and the control constants.

For the second component, we estimated the deformation kernel of a patient-specific 3D lung model extracted from 4D High-Resolution Computed Tomography (HRCT). This component has been previously discussed in detail for the normal lungs [25] and tumor-influenced lungs [26]. We used a Green's Function (GF) based deformation for modeling lung dynamics. The GF was represented as a convolution of the force applied on each node of the 3D lung model and the deformation operator. The force applied on each node represented the air-flow and was based on the distance from the resting surface (due to gravity). For mathematical analysis the GF was represented in a continuous domain. The displacement of every node was computed from subject-specific 4D HRCT and the force applied on each node was estimated for the supine position. A polar-coordinate representation was used to represent each node in the Green's formulation. We expanded the component functions (applied force, deformation and the deformation operator) using Spherical Harmonic (SH) transformations [27]. The SH coefficients of the deformation were expressed as a direct product of the SH coefficients of the applied force and the SH coefficients of the deformation operator's row. Using this framework, the deformable 3D lung model shown in Figure 3 was implemented in OpenGL.

A key aspect of using a physical and physiology-based model is that the resulting 3D deformation accounts for changes in the patient's orientation as well as diaphragm movement. The changes in the diaphragm position were modeled using the SH coefficients to closely simulate the effects of the diaphragm movement.

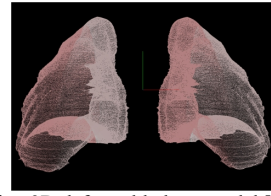

Figure 3. High resolution 3D deformable lung model [28]

### B. DATA DISTRIBUTION SCHEME

To control the 3D lung dynamics remotely, we investigated several alternatives in previous work [7]. We decided to use a bandwidth-conserving method, i.e., we aim at reducing the number of packets sent between nodes, by sending the deformation parameters as a single packet at the start of every breathing cycle. The components of this packet are the input parameters for the lung deformation, which relate the intra-pleural pressure and the lung volume during breathing. Additionally the force applied on each vertex and its elastic properties are also transmitted.

Data is distributed in the form of custom defined data-packages denoted as Control Packet Objects (CPOs). Each packet contains the boundary values of the pressure and volume, control constants, force coefficients and elasticity coefficients.

The boundary values for volume are the lung Functional Residual Capacity (FRC) and the lungs Tidal Volume (TV). "PR" denotes the maximum pressure value while "V" denotes the breathing rate. The control constants $(CP_0,...,CP_N)$ represent the PV relationship required for modeling patient's breathing condition as discussed in the first stage of 3D lung dynamics [6] "N" denotes the number of control constants. The force coefficients $(F_0 - F_M)$ are the SH coefficients that describe the force applied on every node of the 3D model. The number of force coefficients is denoted by "$M$". The elasticity coefficients $(T_0 - T_{2M})$ describe the 3D lung model's elasticity. The deformation of the 3D model at any remote location is computed as the product between the force and the elasticity coefficients.

To compensate for the communication latency, we combined the distributed application with an adaptive delay measurement algorithm [29] which estimates the delay between each pair of users that interact and predicts the next CPO values. Figure 4 illustrates the distribution scheme layers.

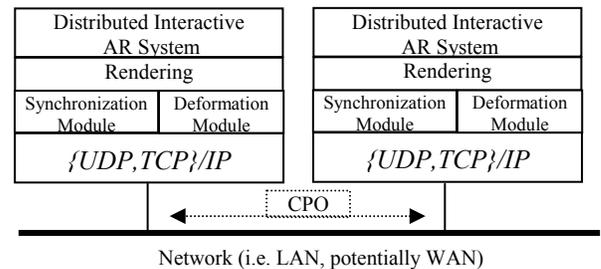

Figure 4. Application Stack Layers

We optimized the data distribution by reducing the number of packets, i.e., for each change in the lung dynamic parameters, a CPO packet was sent. Once the packet is received, each node knows how to drive its own lung deformation simulation.

## VI. RESULTS AND PRELIMINARY ASSESSMENT

We have deployed the prototype using inexpensive Linux based PC's with NVidia™ *GeForce4* graphical processing units on our 100 Mbps local area network using a typical client-server architecture (i.e., the PC connected to the HMD was acting as the server, while other PCs, the clients, on the same network could visualize the 3D model of the Lung remotely). Instead of a real patient we used a HPS from Medical Educational Technologies.

We superimposed the deformable model on the HPS using a Polaris™ infrared optical tracking system. The update cycle is combined with the deformation rendering cycle to obtain an average frame rate of 25 frames per second. Figure 5b shows a camera view of the superimposition of the deformable 3D lung model on the HPS.

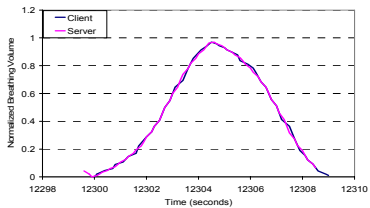

Figure. 5 (a) System components, (b) Augmented View user's view captured with a camera placed behind one eye in the HMD.

### A. VOLUME CONSISTENCY, BREATHING CYCLE

We obtained a smooth and synchronized view between participants, as shown by the normalized breathing volume in Figure 6. The normalized breathing volume reached the same values over time simultaneously at each participant proving that high levels of shared state consistency can be achieved. Interactivity on such a system would be consistent in terms of user's actions.

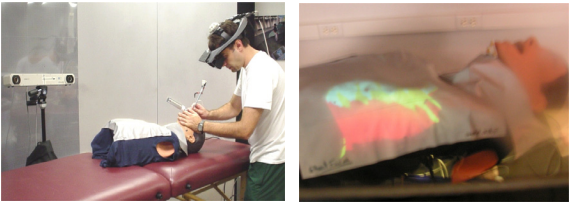

Figure 6. 3D lung model volume as seen at the client and server during one breathing cycle.

### B. PRELIMINARY SCALABILITY ASSESSMENT

To investigate the scalability of the approach, we have increased the number of participants consecutively to two and three. To quantify the scalability of the adaptive synchronization algorithm regarding the number of participants we define a metric analyzing the relationship between the number of participants in the system and the drift values among their views.

Figure 7 illustrates the values of the normalized breathing volume when three participants use the system.

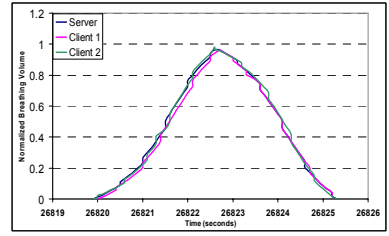

Figure 7. 3D lung model volume as seen by three participants during one breathing cycle.

The average normalized breathing volume drift per breathing cycle was in this case 0.01%. Such small drifts could not be subjectively observed by the users of the system. The shared state consistency ultimately depends on the network infrastructure as well as on the hardware systems complexity. The PC's hardware attributes involved in the three participants setup are described in Table 1.

**Table 1. Hardware system attributes**

| Node no. | CPU (GHz) | RAM (GB) | GPU (GeForce) |
|---|---|---|---|
| 1(server) | 2.8 AMD | 0.5 | Ti4200 |
| 2(client) | 2.4 Intel | 1 | Ti4800 |
| 3(client) | 1.5 AMD | 0.5 | Ti4200 |

## VII. DISCUSSION

Our prototype focuses on coupling physics and physiology-based subject-specific 3D deformable lung models with distributed AR based visualization. The 3D deformable models being generated from 4D HRCT imaging show the effective role that such advanced systems can play in deciding clinical interventions for a wide-range of disease states. Coupled with the distributed AR based visualization, such deformable models facilitate the prototype to have user-specific views of the subject-specific lungs.

From the clinical usage perspective, the prototype can be considered for training and planning under three conditions: (i) without any invasive intervention, (ii) with minimal invasive intervention, and (iii) with thorough invasive intervention.

**(i) Without any invasive intervention:** Such clinical scenarios would include investigating a patient's general breathing patterns and discomfort (e.g. dyspnea). For this case, the prototype allows the user to view subject-specific breathing under different physical conditions and orientations of the patient.

**(ii) With minimal invasive intervention:** Such interventions would include procedures such as intubation, endoscope and needle insertion etc. During such interventions, the proposed AR prototype would be an effective tool since it can show the position of the minimally invasive tools as well as the breathing changes that are caused by the subjective discomfort and the effect of the clinical intervention.

**(iii) With thorough invasive intervention**: Such interventions would include pre-planned procedures such as lung transplants and lung volume reduction. Under such interventions, AR would be an effective tool for visualizing pre-operative conditions and post-operative prognosis for the patient. For instance, in the case of lung transplants, care needs to be taken regarding the changes in the subject's breathing pattern caused by (i) Pleural space complications such as Pneumothorax, (ii) Parenchymal space complications such as Empyema and (iii) Opportunistic infections such as Pneumonitis. Such complications may be avoided by visualizing the patient's breathing morphology in an AR based environment. Simulating intra-operative conditions would heavily rely on the bio-mathematical interactive 3D models that can accurately account for user-induced variations in the subject's anatomy. For a subject-specific lung, developing such 3D models is currently an open research problem.

We are currently in collaboration with the Department of Radiation Oncology at M.D. Anderson Cancer Center for optimizing the prototype for tracking tumor motion and morphological changes during high-precision radiation therapy.

The key technical issue that pertains to the applicability of such a prototype deals with the choice of equipments and the motion compensation. One may note that different tracking systems may be employed along with AR. Of particular importance is the electromagnetic tracking, which is widely used for AR applications. However, care needs to be taken on its usage since the interference between the tracking system and the surgical (metal & carbon fibre) tools present in the room can induce errors in the tracking process. Recently, micro-trackers (approx. 1 mm diameter) have opened the use of magnetic trackers in the surgical room.

From an image perspective, the respiratory motion compensation deals with image blur caused by the imaging system. Since we extract the 3D models using a 4D HRCT imaging system (with breath-hold maneuver) the effect of the image blur in the 3D models is negligible. From an image-streaming perspective, motion compensation issues are generally dealt during image encoding and transmission over a network. However, such image-streaming approach cannot be used in a distributed AR environment as users at two different locations will have two different views of the 3D lungs. We have thus eliminated the image-streaming steps by replacing them with a data distribution scheme that can synchronize the individual views of a group, as the simulation runs on different computers connected through the network.

From a simulation and visualization perspective, the respiratory motion compensation refers to non-smooth simulations and can be considered under the following three different conditions:

**(i) Respiratory motion compensation without patient's motion and breathing changes:** In [5] we have presented a detailed account of the respiratory motion and the generation of real-time physically-based 3D deformable lung models. A key aspect of the 3D lung deformations is their ability to satisfy real-time constraints. Using state-of-art graphics processing units, we display the lung deformation at a rate of 75 frames per second (without tracking).

**(ii) Respiratory motion compensation with patient's breathing variations:** In recent work, we have accounted for the changes in the lung dynamics caused by the changes in the physiological (caused by lung tissue degenerations) [26] and behavioral conditions (caused by subjective perception of discomfort) [30] of the patient. The tissue degenerations were accounted by modifying the deformation kernel, and the subjective perception of the discomfort was simulated by modifying the PV curve.

**(iii) Respiratory motion compensation with patient's motion:** One may note that the task of simulating respiratory motion with patient's motion compensation is a complex task. In our approach we have compensated for the patient's motion by modifying the SH coefficients of the applied force by interpolating among a set of pre-computed applied forces for each orientation of the subject. The interpolation did not induce any computational delay and did not affect the real-time nature of the simulation.

## VIII. CONCLUSIONS AND FUTURE WORK

We have presented the integration of a distributed interactive planning prototype that uses a deformable lung model combined with the AR capabilities. The distribution of deformable 3D models at remote locations allows efficient communication of concepts and generates a vast potential for collaboration and training.

An important assumption in the experiments is that the distributed system is composed of fairly homogeneous nodes i.e., each PC has approximately similar rendering and communication capabilities. The experiments were performed on a low latency network with up to three nodes, one server and two clients. We plan to increase the number of users to further test the scalability of the prototype and to assess the efficiency of the system from the human factors perspective.

Future work involves adding disease states such as the chronic obstructive pulmonary disease, dyspneac breathing, and pneumothorax influenced 3D lung deformations.

### ACKNOWLEDGMENTS

We wish to thank our sponsors: the Link Foundation, the Florida Photonics Center of Excellence, the US Army STRICOM and NVIS inc. for their support.

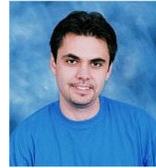
**Felix G. Hamza-Lup, Ph.D** is Assistant Professor of Computer Science at the School of Computing at the Armstrong Atlantic State University, Savannah, Georgia. He received his B.Sc. in Computer Science from the Technical University of Cluj-Napoca, Romania his M.S and Ph.D. in Computer Science from the University of Central Florida. His current focus is the development of novel human computer interaction techniques based on augmented reality paradigms. He is a member of the ACM, IEEE, and the Upsilon Pi Epsilon.

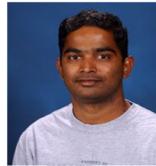
**Anand P. Santhanam, M.S.** is doctoral candidate at the School of Computer Science at the University of Central Florida. He received his B.E degree in Computer Science from the University of Madras, India, and his M.S degree in Computer Science from the University of Texas, Dallas in 2001. His current research interests include real-time graphics and animation in virtual environments. Anand has been awarded the Link Foundation Fellowship in Advanced Simulation and Training at the University of Central Florida.

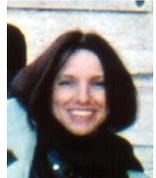
**Celina Imielińska, Ph.D.** is Associate Research Scientist affiliated with Columbia University College of Physicians and Surgeons Office Scholarly Resources and Department of Medical Informatics, and Department of Computer Science. She has a B.E. degree in Electrical Engineering from Politechnika Gdanska, in Gdansk, Poland; and a M.S. and Ph.D. degree in Computer Science from Rutgers University, in New Brunswick. Her current interests are medical imaging, 3D visualization, and computational geometry.

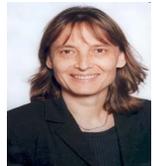
**Jannick P. Rolland, Ph.D** is Associate Professor of Optics, Computer Science, Electrical Engineering, and Modeling and Simulation at the University of Central Florida. She received a Diploma from the Ecole Superieure D'Optique in Orsay, France, in 1984, and her Ph.D. in Optical Science from the University of Arizona in 1990. Dr. Rolland is Associate Editor of Presence (MIT Press), and has been Associated Editor of Optical Engineering 1999-2004. She is the UCF Distinguished Professor of year 2001 for the UCF Centers and Institutes, a member of IEEE, and a Fellow of the OSA.

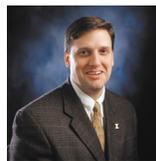
**Sanford L. Meeks, Ph.D.** is Director of Radiation Physics at M. D. Anderson Cancer Center Orlando. He received his M.S. in physics from Florida State University and his Ph.D. in Medical Physics from the University of Florida in 1994. Previously, he was Associate Professor and Director of Radiation Physics at the University of Iowa. He has co-authored 15 book chapters and more than 50 peer-reviewed publications related to stereotactic radiosurgery, intensity modulated radiation therapy, and image-guidance.